# A new multi line-cusp magnetic field plasma device (MPD) with variable magnetic field for fundamental plasma studies

A. D. Patel, M. Sharma, N. Ramasubramanian, R. Ganesh, P. K. Chattopadhyay

*Institute for Plasma Research, HBNI, Bhat, Gandhinagar, Gujarat - 382428, India*

**Abstract:**

One of the fundamental problems is the understanding of physics of electrostatic and electromagnetic fluctuations in multi-scale plasma turbulence. Especially so, in continuously connected plasma regions with varying degree of magnetization. Examples range from multiscale plasmas in Magnetron-like devices to astrophysical plasmas confined by magnetic dipole structures, solar wind driven collision-less and weakly collisional plasmas around Earth, to mention a few. Such plasmas are dominated by both electron scale and ion scale physics as well as finite beta effects. To investigate such processes in laboratory experiments requires excellent control of continuously connected regions of nearly zero plasma beta with finite beta regions as well as the gradient length scales of mean density and temperatures. To address some of these phenomena at laboratory scale, a new multi-line cusp configured plasma device (MPD) consisting of electromagnets with core material has been constructed with a capability to experimentally control the relative volume fractions of magnetized to unmagntized plasma volume as well as accurate control on the gradient length scales of mean density and temperature profiles. The hot tungsten cathode produced Argon plasma in the MPD has been characterised using single Langmuir probes. Argon plasma has been produced in the device over a wide range of pressure $5 \times 10^{-5}$ mBar to $1 \times 10^{-3}$ mBar, achieving plasma density range from $10^9$ to $10^{11}$ cm$^{-3}$ and temperature in the range 1eV to 8eV. The plasma profiles thus measured radially along the non-cusp region (in between the magnet) show a uniform plasma region across the axis for about 8-10 cm diameters, where the magnetic field is very low such that the ions are unmagnetized. Beyond that region, the plasma species are magnetized and the profiles show gradients both in temperature and density. The gradient in magnetic field can be controlled by the current in the electromagnet. The electrostatic fluctuations measured using Langmuir probe radially along the non-cusp region shows less than 1% ($\delta I_{isat}/I_{isat} < 1\%$). The total volume of the plasma is $1.2 \times 10^5$ cm$^3$ and 10% of plasma volume is quiescent in nearly field free region and this volume is controllable by changing magnetic field. These features open up several new and hitherto unexplored physics parameter space relevant to both laboratory multiscale plasmas as well as astrophysical plasmas.

**Introduction**

Confined plasmas have a broad range of applications from the laboratory to industry and to the realization of controlled fusion. Conventionally magnetic fields have been used to confining the plasma though with different configurations, but plasma confinement is still an open problem in general. Some of the configurations studied were linear devices like mirror to cusp geometries and toroidal devices like stellarators, tokamaks etc. Of particular interest in the multi-cusp configuration is the nearly (B~0) null field in the centre of the confining region, while the finite value of field in the edge regions confining the plasma well. In this configuration the centre of the curvature of the confining magnetic field will not be in the region of confined plasma. This helps to avoid many instabilities which are usually observed in the laboratory devices and hence high beta ($\beta_e = 1$, the ratio of plasma pressure to magnetic pressure) plasmas can be confined [1].

Multi-cusp magnetic field has found wide applications in the development of ion sources [2-4], plasma-etching reactors [5] etc. The surface cusp magnetic field can able to confine quiescent plasma with densities up to 100 times larger than in the double plasma device [6]. For example, the negative ion source with multi-cusp configuration has demonstrated its capability for high current low emittance and stable H$^-$ ion beams which is essential for the new generation of accelerators [7]. Also, these multi-cusp geometries are being used for basic plasma-physics studies in laboratory devices due to their ability to confine large-volume uniform quiescent plasmas [6]. For example, because of this condition (B~0) in the centre, drift wave oscillations as observed in the classical Q-machines are not triggered [8]. But the discontinuities in the confining field in the boundary leak the plasma through the cusp regions. The exact scaling for the amount of plasma leaking is not known, though many scaling exists ranging from ion gyro-radius to electron gyro-radius apart from the so-called 'hybrid gyro-radius' [10-14]. Also the various fluctuations present in the edge regions (low Beta) and their effect on the central high beta region have not been studied before. It is because most of the devices made were with very small volume low beta region compared to the volume of the high beta region.

Even so, the plasma confined in the central region of the multi-cusp field is very useful to study the various fundamental plasma phenomenon such as landau damping, nonlinear coherent structure, wave –particle trapping, and un-trapping etc., requires reasonably quiescent plasma. The ratio of $m_e/m_i$ crucially controls the finite beta effect. Thus a control on $m_i$ becomes necessary. Nonlinear ion frequency mode studies require variable mass species. Thus it becomes necessary to have a flexible plasma source. Existing quiescent machine (Q-machine)typically use contact ionization plasma. For the above said reason we have developed a hot cathode based plasma source capable of several studies mentioned above.

Most of the earlier studies in the multi-cusp field configuration employed either permanent broken line magnets [2-4] or electromagnets without any profiling of the field. The former case with permanent magnets with fixed field values, it was not possible to do various magnetic field scaling studies. Simple electromagnets might also give unwanted edge effects.These devices do not have control over quiescent plasma volume and finite beta volume. To have these characteristics simultaneously i.e. variable field values and uniformly profiled field, controllable quiescent plasma volume and finite beta volume it is thought here, that a multi-line cusp configuration using electromagnets with profiled core material. A new device has been built with these characteristics has been described here.The basic characterization of plasma has been carried out under different magnetic field strengths. This paper presents the details about the device and the initial results and inferences from the basic characterisation.

The construction of the device is described in Section II, Section III contains the experimental results and discussions and the conclusion of the observed experimental results will be discussed in Section IV.

I. **Experimental setup**

Assuming typical Argon plasma with 2-8eV electron temperature, the field values and geometries were worked outusing FEMM (Finite Element Method Magnetic) [9] simulation. It is found that with six magnets kept in the multi-cusp configuration at 60°each along the circumference of a 40 cm diameter cylinder, a finite plasma region (10cm diameter and length 60 cm) where the ions are not magnetized (high beta), can be achieved.This will also allow having a finite volume of low beta region. The schematic of the complete device is shown in the figure 1. The device consists of a vacuum chamber with pumping solutions, magnet system, a filament based source for the plasma production, and basic probe diagnostics.

A. **Vacuum Chamber**

The main chamber of the experimental system is shown in figure 1. It is made up of non-magnetic stainless steel (SS-304). The main cylinder chamber is 40 cm diameter and 150 cm length, with a wall thickness of 1cm.  This chamber is pumped out by a Turbo-Molecular Pump (430 litres/sec) backed by a rotary pump through a conical reducer on one side. The other side of the chamber is closed with a 400 CF flange containing view ports and vacuum feedthrough for electric probes.The whole system can be pumped to a base pressure of 1 x $10^{-6}$ mBar. The vacuum gauges are connected at the middle of the chamber for the measuring the vacuum. The chamber has many radial ports with provisions for feedthroughs of electrical connections to the source inside, for viewing, and some ports for movable probe diagnostics. The six electromagnets are firmly supported to the main chamber by a non-magnetic C-clamp.

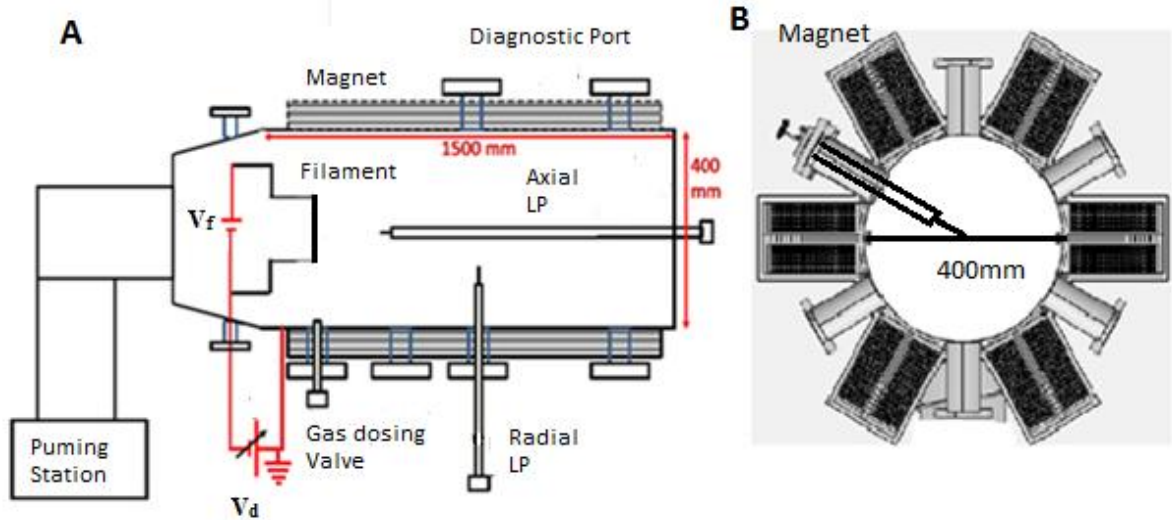

FIG. 1: Schematic diagram of (A) Experimental setup and (B) Chamber End cross sectional view of the multi-line cuspmagnetic field plasma device

### B. Magnetsystem

It consists of six electromagnets in which each one is made with double pan-cake windings using hollow copper pipes, cross-section of which is shown in the figure 2. The hollow copper pipes are used for forced water cooling and the pan-cake windings help for effective cooling using parallel lines. The physical dimensions of the rectangular magnet are in centimetres 132, 19.5 and 14 respectively for length, width and height. A core material having dimensions in centimetres 120, 2 and 12 respectively for length, width and height is placed in each of the magnets so as to compare the plasma characteristics with the experiments done before with permanent magnets. This core material is made up of Vacoflux-50 which is an alloy of iron and cobalt in the ratio 50:50.To make the field uniform and to avoid the edge effects, the core material was smoothened by making a curved face at the edge. After profiling the material, a high temperature annealing was done to strengthen the edges. The mapping of the magnetic field was done using triple axis gauss probe. The measurements of the field for a given current were done at many locations as well as different plane of the device. Figure 3 shows the simulated (using FEMM) 2-D colourcontour plot of the field on the r-theta plane at the centre of the chamber. The magnetic field is measured in cusp region along the magnet and in the non-cusp region which is in between two consecutive magnets as shown in figure 3. The comparison between measured and simulated magnetic field when magnets are energised with a current of 100A and 150A is shown in figure 4, shows good agreement (matched more than 90%). The slight difference in the values between the measured and simulated beyond 12cm is due to the finite boundary conditions assumed in the simulation to reduce the computational times.This set of magnets provide the control over the number of null regions to be created inside the plasma by energizing the required magnets only. Also by changing the current in each magnet, the positions of the null regions can be adjusted as shown in the figure 3(b) and 3(c).

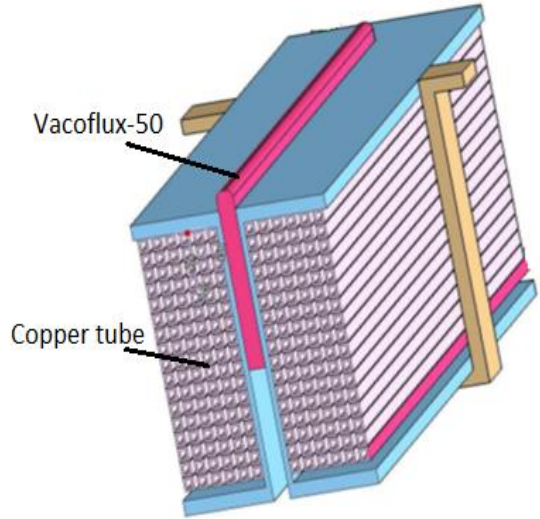

FIG. 2: Cross-sectional of the electromagnet showing copper pipes and core material vacoflux-50.

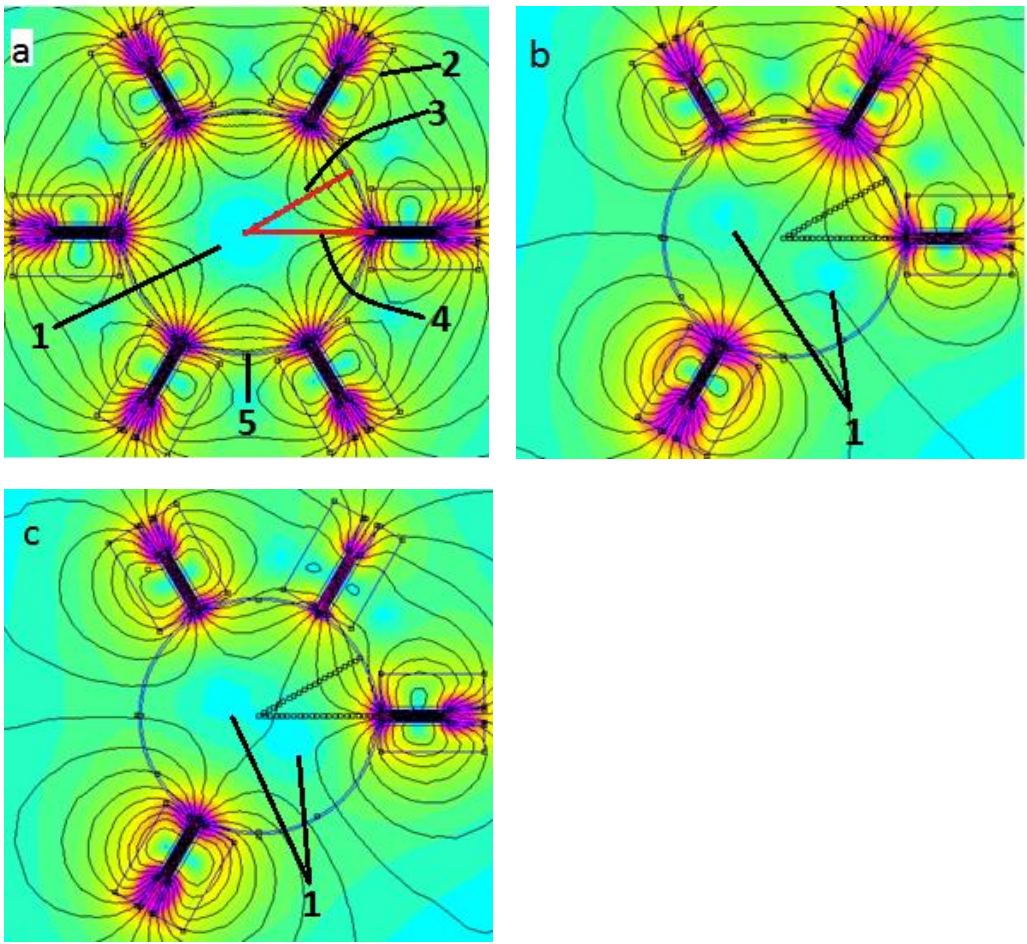

FIG 3: (a) Contour plot of the field lines in (r, θ) plane from FEMM simulation along with the chamber and the position of the magnets; (1) magnetic null region, (2) magnet, (3) non-cusp region, (4) cusp region and (5) chamber wall. The radial profile of all plasma parameters have been measured along the non-cusp region until unless mention specifically. (b)

Contour plot showing two null regions and (c) two null region are closer to each other which shows that the magnet system has a flexibility of different configurations.

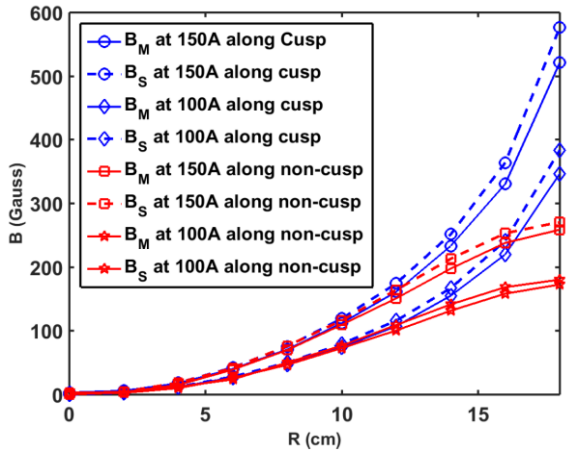

FIG. 4: Magnetic field (both simulated and measured) along the radius through the cusp and non-cusp regions at two different currents 100 A and 150 A

### C. Plasma Production

Hot cathode plasmas made of tungsten filaments are generally quiescent [6]. Studying fundamental plasma phenomena such as landau damping, nonlinear coherent structure, wave –particle trapping and un-trapping etc., require a reasonably quiescent plasma while the field free region is an added advantage. The ratio of $m_e/m_i$ crucially controls the finite beta effects. Hence it is necessary to have a source to study with different masses of ions, which requires the source to be compatible with different gaseslike Argon, Krypton and Xenon etc. With these requirements, a filament based source has been designed and fabricated. It is a two dimensional vertical array of five filaments each having length 8cm made using 0.5mm diameter tungsten wires. This source is fitted from the conical reducer such that the filaments areinside the main chamber itself where the magnetic field is low. Also it has been taken care to push the source well inside the main chamber to keep it away from the edge of the magnets. These filaments are powered by a 500A, 15V floating power supply while it is normally operated at around 16-19A per filament. The chamber was filled with Argon gas through a needle valve to a pressure range between 5 x $10^{-5}$ to 1 x $10^{-3}$ mBar. The source is biased with a voltage of -76 V with respect to the grounded chamber walls using discharge power supply. The primary electrons emitted from the filaments travel in the electrical field directions, while they are confined by the cusp magnetic field lines. Because of mirror effects due to cusp configuration, electrons will move back and forth between the poles. Theseelectrons collide with background Argon atoms and hence produce plasma by electron impact ionization. Figure 5 shows the photograph of the Argon plasma taken through the view port attached to the end flange of the chamber. It can be observed that the plasma has having sharp wingslike stars along the cusp regions as shown in figure 5.This is due to the electrons traversing the cusp region multiple times and making more collisions than that of the non-cusp region. The uniform sharpness in all the wings shows the presence of the uniform magnetic field due to the electromagnets with the core material.Also since the azimuthal width of the wings are representative of the cusp leak width, it can be seen that the width of the wings is decreasing radially as the field is increasing.

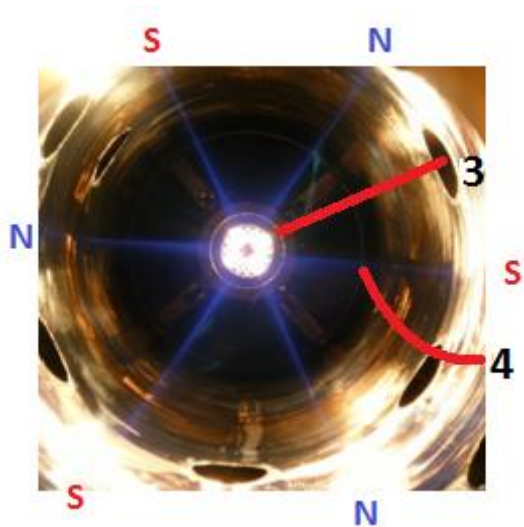

FIG. 5: Photograph of Argon plasma taken from the viewport fitted at end of the chamber in multi cusp plasma device.

### D.     Basic Diagnostics

To characterize the plasma, Langmuir probes are used. The data from the probes are used to estimate the plasma parameters like the mean plasma densities, mean electron temperatures and the fluctuations. For measuring the plasma density and electron temperature, a sweeping circuit (as shown in the figure 6) of frequency 1Hz and voltage sweep from -80V to +30V is used. An analysis code has been developed in MATLAB for quick processing of the data. Figure 7 generated from the MATLAB code show the process of analysis the Langmuir probe *I-V* characteristic of plasma. A typical *I-V* characteristic of the probe data is shown in figure 7(A).

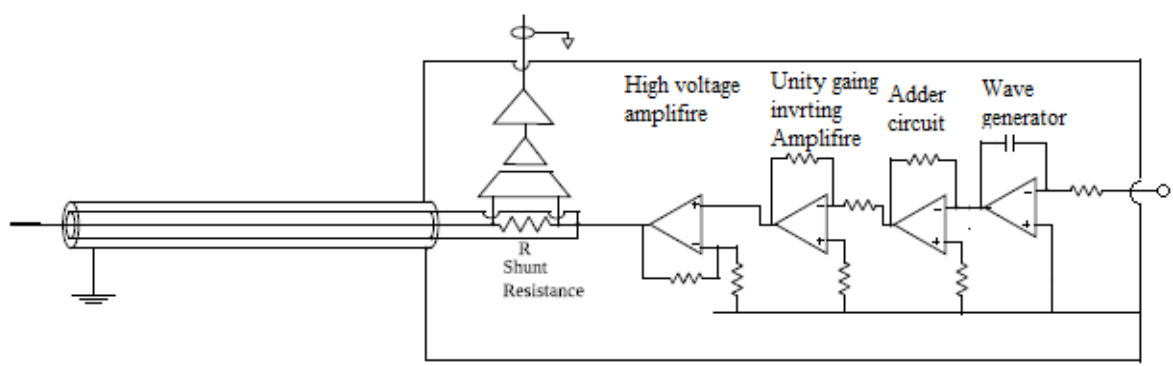

FIG. 6: Schematic of Langmuir probe circuit diagram used for acquiring *I-V* characteristics of plasma

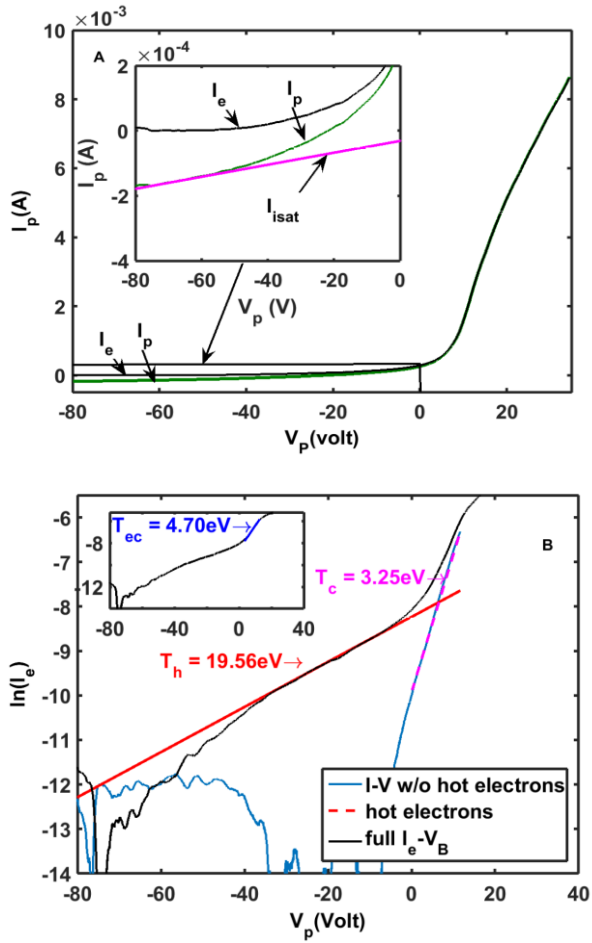

FIG. 7: Plots show the process of analysis for measured Langmuir probe *I-V* analysis. In figure (A) green line shows measured Langmuir probe *I-V* characteristic of plasma, magenta line shows fitted ion saturation current ($I_{isat}$) and black line shows the electron current ($I_e$) after subtracting ion saturation current. FIG. B show variation of $ln(I_e)$ with probe potential ($V_P$), dashedmagenta line shows the cold plasma electron temperature after subtracting hot electron current, red line shows the hot electron temperature. Subfigure in figure (B) shows electron temperature without eliminating hot electron effect (dark blue line).

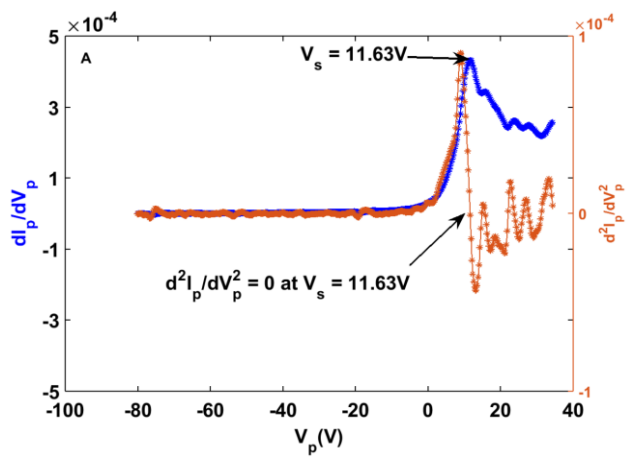

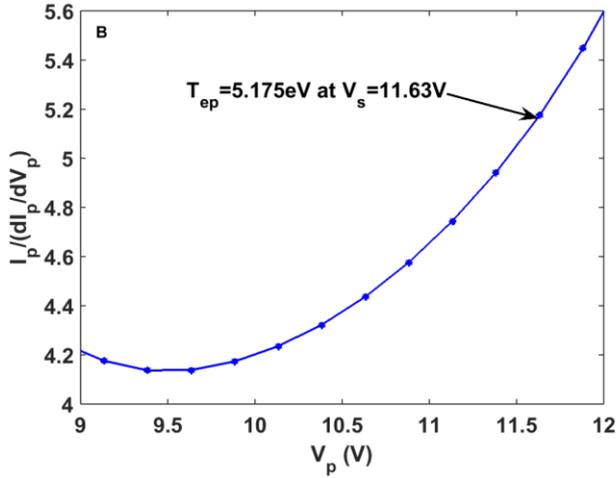

FIG. 8: Figure (A) shows variation of first and second derivative of measured *I-V* characteristic with probe potential ($V_P$), Figure (B) shows variation of the ratio of probe current to first derivative of probe current with probe potential ($V_p$).

### 1. Analysis of electron temperature

For bi-Maxwellian plasma, there is two electron temperature $T_h$ and $T_{ec}$. Based on kinetic definition of temperature,

$$T_e = \frac{1}{3} m_e \int_{-\infty}^{\infty} v^2 f(v) dv \quad \ldots\ldots\ldots\ldots\ldots\ldots\ldots\ldots\ldots (1).$$

An effective temperature can be derived by assuming electron energy distribution function (EEDF) bi-Maxwellian plasma ($f(v) = \alpha f_h(v) + (1-\alpha) f_c(v)$), Where $f_h(v)$ is the hot electron distribution function and $f_c(v)$ is the cold electron distribution function and $\alpha$ is the fraction of electron in the hot electron tail is equal to the current at the plasma potential due to the hot electron divided by the electron saturation current [15, 16],

$$T_{eff} = (1-\alpha) T_{ec} + \alpha T_h \quad \ldots\ldots\ldots\ldots\ldots\ldots\ldots (2).$$

Where $T_{ec}$ and $T_h$ are cold and hot electron temperature respectively.

A measured Langmuir probe *I-V* characteristic is shown in figure 7(A) as a green line. To measure the effective electron temperature ($T_{eff}$), first the digital noise has been removed by smoothing the Langmuir probe *I-V* characteristic, and fitted the linear line at higher negative voltage side to measure the ion saturation current ($I_{isat}$) and substrate it from the probe current $I_p$ it give us electron current $I_e$ as shown in figure 7(A). To estimate the hot electron temperature ($T_h$) from the ln($I_e$) Vs $V_P$ straight line has been fitted at the linear region of tail part of the measured *I-V* as shown in figure 7(B) as red line, using this $T_h$, we can find the hot electron current. After subtracting it from the electron current $I_e$, we can estimate the cold electron temperature ($T_{ec}$) by taking the ln($I_e$) Vs $V_P$, and fitted a line at the Maxwellian plasma region of *I-V* as shown in figure 7(B) as magenta dash line [15, 16]. The subfigure of figure 7(B) shows the electron temperature as dark blue line without subtracting hot electron. The ratio of hot electron current to electron saturation current at plasma potential ($V_s$) gives the value of $\alpha$. For plasma potential measurement a spline fitting has been used on probe *I-V* characteristic and take the first derivative and second derivative of probe current $I_p$ with respect to probe potential $V_P$ as shown in figure 8(A). Using the value of $\alpha$, hot electron temperature ($T_h$) and cold electron temperature ($T_{ec}$) we can calculate effective electron temperature ($T_{eff}$) from equation 2.

The ratio of probe current to first derivative of probe current ($I_p$) with respect to probe voltage ($V_P$) at the plasma potential ($V_s$) gives also the electron temperature ($T_{ep}$) [17],

$$\left. \frac{I_p}{dI_p/dV_p} \right| = T_{ep} \quad \ldots\ldots\ldots\ldots\ldots\ldots\ldots\ldots (3).$$

We notice that effective electron temperature ($T_{eff}$) and electron temperature ($T_{ep}$) from equation 2 and 3 respectively are nearly same within error bar.

### 2. Analysis of plasma density

In a collision less single ion species plasma the ion saturation current to the probe is expressed by

$$I_{isat} = \gamma e\, n\, A_P \sqrt{\frac{k_B T_e}{m_i}} \quad \text{........................} \quad (4)$$

Where $n$ is the ion density, $A_p$ is the probe area, $m_i$ is the mass of the ion. The constant $\gamma$ depends on the probe sheath thickness. For $\xi = r_p/\lambda_D > 3$, where $r_p$ is probe radius and $\lambda_D$ is Debye length,

$$\gamma = \zeta \chi^\beta / 4 \quad \text{...........................} \quad (5)$$

Where $\chi = \dfrac{e(V_s - V_p)}{k_B T_h}$ and $\zeta = 1.37 \xi^{0.06}$ and $\beta = \xi^{0.52}$ [19].

The equation of $I_{isat}$ for $\xi = r_p/\lambda_D \leq 3$ is given by OML theory [20],

$$I_{isat} = 1.13 e\, n\, A_P\, \chi^{0.5} \sqrt{\frac{k_B T_e}{2\pi m_i}} \quad \text{....................} \quad (6)$$

We calculate the densities using equation (4) by changing different electron temperature as describe above, we notice that the densities are also nearly same with in error bar. For error calculation in density, equation $\dfrac{\delta n}{n} = \sqrt{\dfrac{\delta I_{isat}}{I_{isat}} + \dfrac{1}{2}\dfrac{\delta T_e}{T_e} + \left(\dfrac{\delta n}{n}\right)_s}$, is considered, where $\dfrac{\delta I_{isat}}{I_{isat}}$ error in ion saturation current, $\dfrac{\delta T_e}{T_e}$ error in electron temperature measurement, $\left(\dfrac{\delta n}{n}\right)_s$ is statistical variation in density. For the rest of article, the densities estimated using equation (4) and effective electron temperature ($T_{eff}$) are only considered.

## II. Results and discussions

The Argon plasma thus produced has been characterized by measuring the radial profiles of plasma parameters and its fluctuation along the non-cusp region for different magnetic field values. The details are as follows:

### a. Profiles of mean density and mean electron temperature

The radial profiles of plasma density and electron temperature are shown in figure 9 and 10 at different magnetic field values along the non-cusp region with Argon gas background pressure is $2 \times 10^{-4}$ mbar in the middle plane of the device which is 60cm apart from the source. This shows that for a cylindrical region of about 10 cm centred along the axis is having uniform plasma. From the radial electron temperature profile, it is observed that the field free region is hotter up to 5cm and after that temperature decreases radially. This is due to the primary electrons along the axis rush to the other edge with very less collisions because of very low magnetic field. But then the field increases, though slightly, the electrons are magnetized and start following the magnetic field and hence heating all the neutral and ions on the way. Beyond 5 cm, the number of collisions are too high, hence the temperature starts decreasing.

Figure 11 shows the variation of plasma density and electron temperature at the centre of the device (R=0) with different field values for a background pressure of $2 \times 10^{-4}$ mbar. It can be observed that the density increases with the higher magnetic field values while the temperature decreases slightly. With higher field values, the confinement of primary electrons is increased. This increase in confinement effectively makes more collisions and ionizations which in turn increases the density. For a given energy, when the density increases, the average energy per particle decreases and hence the temperature decreases as it is observed.

Figure 12 shows the radial variation of electron plasma beta ($\beta_e$, plasma pressure to magnetic field pressure) along the non-cusp region with Argon gas background pressure is 2 x $10^{-4}$ mbar in the middle plane of the device. At the centre of the device (R=0) plasma beta is maximum nearly one after that it decreases with radial distance. The figure shows nearly zero beta region comparable in volume with finite beta region. Finite beta region enjoys the benefit of absolute magnetohydrodynamic stability and nearly zero beta region enjoy the benefit of electrostatic stability.

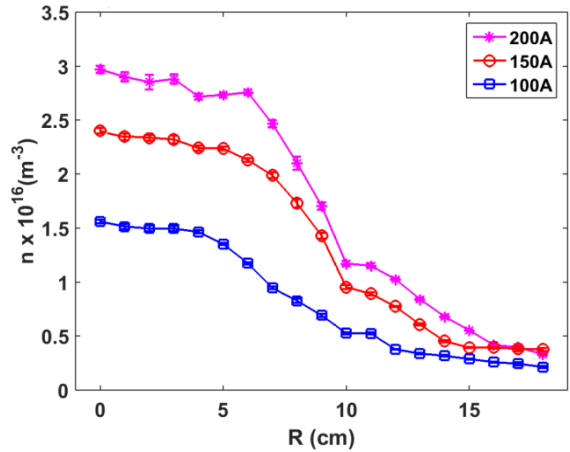

FIG. 9: Radial profiles of plasma density along the non-cusp region at magnets are energies with 100, 150, and 200A current, Argon gas background pressure is 2 x $10^{-4}$mBar.

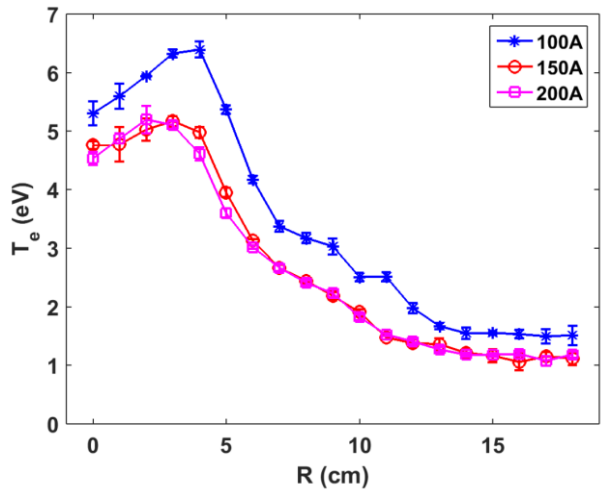

FIG. 10: Radial profiles of electron temperature along the non-cusp region at magnets are energies with 100, 150, and 200A current, Argon gas background pressure is 2 x $10^{-4}$mBar.

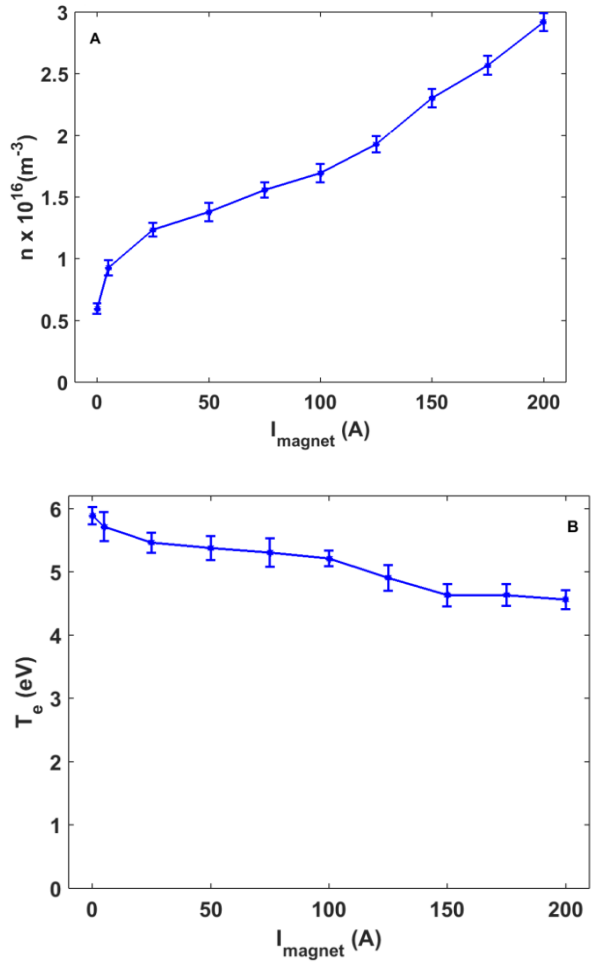

FIG. 11: Show the variation of plasma density (Figure A) and electron temperature (Figure B) with magnets is energies with different current at the middle point of the device (R=0cm), Argon gas background pressure is 2 x $10^{-4}$mBar.

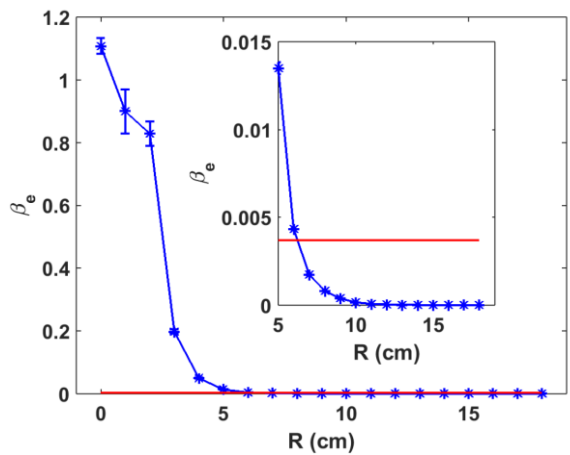

FIG. 12: Radial profile of plasma beta($\beta_e$) in non cusp region whenmagnets are energies with 150A current and Argon gas background pressure is 2 x $10^{-4}$mBar. The red line shows the value of plasma beta line $y = \sqrt{m_e/m_i}$ for reference.

### a. *Quiescence level*

Figure 13 shows the typical fluctuations of the ion saturation current at different radial location as measured using the ion saturation current in voltage across 10 kΩ resistance when probe is biased at -80V. It can be seen that the fluctuations are increasing radially outward. Figure 14 shows the variation of fluctuations level along the radial in the non-cusp region. It can be seen that the ion saturation fluctuations are ($\delta I_{isat}/I_{isat}$) < 1% up to a distance of R=6cm which is a signature of the quiescent state. It is to be noticed that the fluctuation levels are very low even though there are some gradients in temperature in that low field region (upto R=6cm). This cylindrical region with uniform density and quiescent plasma can be used to study for different fundamental phenomena. After R=6cm, the fluctuation level increases, there is a sharp jump around R=10cm and it reached almost 11% at R=15cm as it can be seen from the figure 14. The more detailed analysis of these fluctuation studies will be published elsewhere.

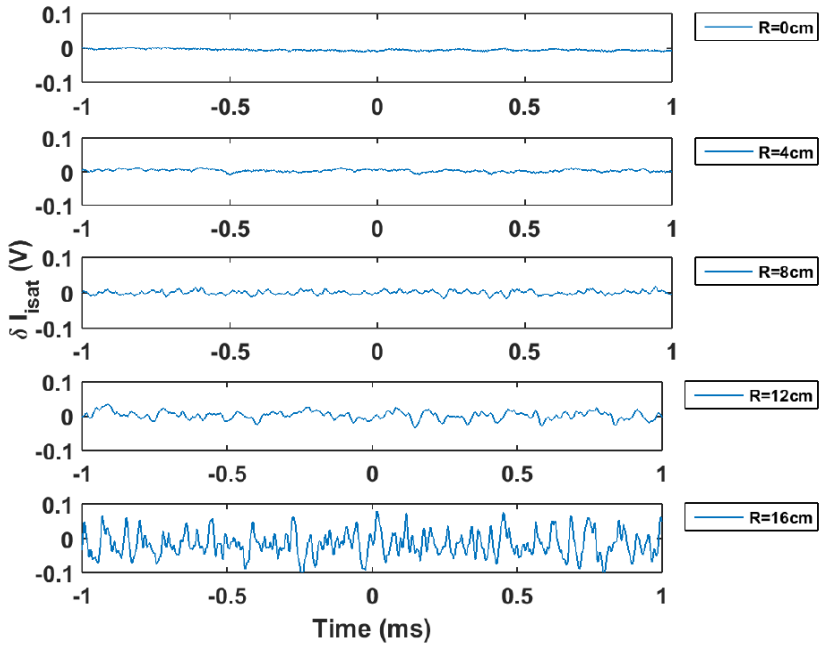

FIG 13: Time profiles of ion saturation current fluctuation in voltage across 10kΩ resistance when probe is biased with -80V at different radial location along the non-cusp region when magnets are energising with 150A current and Argon gas background pressure is 2 x $10^{-4}$mBar.

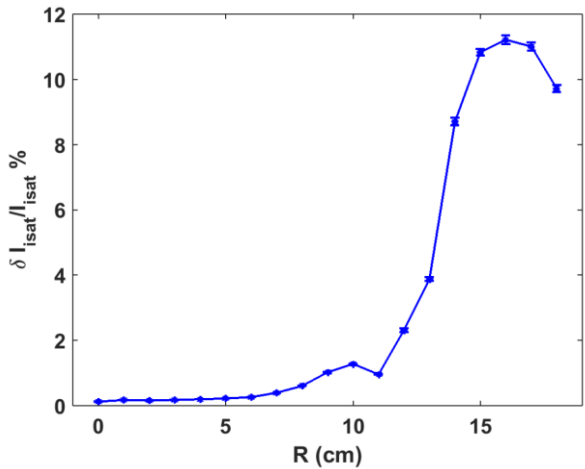

FIG. 14: Variation of fluctuation level of ionsaturation current radially along the non cusp region when magnets are enegized with 150A current and Argon gas background pressure is 2 x $10^{-4}$mBar.

III. **Conclusion**

A new multi- line cusp plasma device has been design and constructed in Institute for Plasma Research (IPR). The hot cathode filament based source placed in low magnetic field region of multi cusp magnetic field.The uniqueness of the device is the profiled core of the electromagnets which will allow changing the field values with the effect of using different permanent magnets.The initial characterisation with Argon plasma with changing magnetic field values has been performed. The radial profile of the plasma parameters has been measured in the middle plane of the device along the non-cusp region. The plasma is found to be very quiescent($\delta I_{isat}/I_{isat}$) < 1%) up to 6 cm radius from the axis as measured from the fluctuations of ion saturation current. After 6 cm, the fluctuation levels are found to be increase. The plasma density is uniform over a 10cm diameter of the chamber. The device has a nearly zero beta region comparable in volume with finite beta region has been kept to enable fluctuation studies of both regions simultaneously. The device operates over a wide range of pressure 5 x $10^{-5}$mBar to 1 x $10^{-3}$mBar and different discharge voltage ($V_d$~ -40 to -100V) which has control over collisionality and energetic electron. This device can also able to produce Helium, Neon, Krypton, Xenon and their different combination of plasma. This machine is also design for accommodating many plasma source like oxide coated cathode [21], hot tungsten plate based contact ionization source for alkali plasma [22] etc. This device is also designto operate with source at the both end.All these above functional flexibility and wide range of operation regimes make this device for carrying out study fundamental plasma phenomena such as landau damping, nonlinear coherent structure, wave–particle trapping and un-trapping, plasma waves, and nonlinear ion frequency mode studies with variable mass species etc. The study performed in this manuscript also useful in ion sources, thin film technology, plasma vapour deposition, plasma etching reactors etc. The detailed analysis of fluctuation and leak width (the width from which plasma is loss at the pole [3]), and also effect of magnetic field strength on thesewill be studied and published elsewhere. Also new experiments are being planned like, plasma wave studies, wave-particle interaction and change the design of source to understand various fundamental plasma physics with this new device.

Acknowledgement

The authors would like to thanks to Professor P. K. Kaw and Professor Y. C. Saxena for numerous fruitful discussion.